\documentstyle[aps,epsfig,multicol,eqsecnum]{revtex}    

\begin{document}

\title{A Fast Algorithm for Generating Long Self-Affine Profiles}
\author{Ingve Simonsen$^{a,b,}$\footnote{Email: Ingve.Simonsen@phys.ntnu.no}
  and 
Alex Hansen$^{b,c,}$\footnote{Email: Alex.Hansen@phys.ntnu.no}}

\address{$^a$Department of Physics and Astronomy
and Institute for Surface and Interface Science,\\
University of California, Irvine, CA 92697, U.S.A.}
\address{$^b$Department of Physics, Theoretical Physics Group, \\
The Norwegian University of Science and Technology, \\
N--7491 Trondheim, Norway}
\address{$^c$International Centre for Condensed Matter Physics,\\
University of Bras{\'\i}lia, CP 04513, 70919--970, Bras{\'\i}lia, Brazil}

\date{\today}
\maketitle

\begin{abstract}
    We introduce a fast algorithm for generating long self-affine
    profiles. The algorithm, which is based on the fast wavelet
    transform, is faster than the conventional Fourier filtering
    algorithm. In addition to increased performance for large systems, the
    algorithm, named the wavelet filtering algorithm, a priori gives
    rise to profiles for which the long-range correlation extends
    throughout the entire system independently of the
    length scale.
\end{abstract}

\vspace{5mm}
\begin{multicols}{2}
\narrowtext

\section{Introduction}
With the advent of the computer as a serious research tool, there 
has been a revolution in the quantitative description of processes and 
structures that before were deemed too complex.  Two of the key concepts 
used for this description are the fractal and its close relative, the
self-affine structure \cite{Feder}.  In the early eighties, much effort 
was spent identifying and describing various physical systems having
fractal or self-affine structure.  As time went by, focus slowly shifted 
from pure description to asking why such structures would appear.  This   
lead to the development of the science of complex growth phenomena.  Now, 
many aspects of these are well understood.  However, there are still hosts
of interesting but unanswered questions lingering on --- see e.g. Refs.\
\cite{Barabasi,Meakin,Meakin98} for recent reviews.  More recently, focus
has again begun to shift somewhat, and one sees work dealing with the
physical consequences of the presence of fractal or self-affine structures.
A concrete example of these three levels of development may be found in
the study of fracture surfaces.  In the early eighties, Mandelbrot et al.\
\cite{Mandelbrot} characterized fracture surfaces as self-affine, in 
the early nineties attempts were made to understand why fracture
surfaces are self-affine \cite{Bouchaud}.  Recently, phenomena
such as two-phase flow in fracture joints have been studied \cite{Auradou}.

In order to study the physical consequences of the presence of self-affine 
surfaces, algorithms generating these must be found.  There are already
several in existence, see e.g.\ Feder \cite{Feder}.  However, subtle
phenomena require the generation of huge surfaces.  Two aspects of
the algorithms then become important: (i) how self-affine are the surfaces
that are generated, and (ii) how fast is the algorithm.

The most popular algorithm used today is the Fourier filtering 
algorithm. This algorithm, which has a fast implementation thanks 
to the fast Fourier transform, consists of generating in the Fourier 
domain, uncorrelated Gaussian random numbers which are filtered by a 
decaying power-law filter of exponent $-2H-1$ where $H$ is the Hurst
exponent (to be defined in Section \ref{Sec:2}). By taking advantage of 
the inverse fast Fourier  transform, self-affine surfaces in real space, 
with the desirable correlations, are generated.

It has previously been shown \cite{Peng,Prakash} that the Fourier
filtering algorithm has the disadvantage that the self-affine
correlations in the limit of large systems only exists over a fraction
of the total system size.  This is due to aliasing effects.  For large
enough systems this fraction might be well below 1\%.  One might
overcome the above problem by e.g.\ temporarily generating a much
larger surface than actually needed, and only use a small fraction of
the total size.  This is, however, not a very appealing approach, as
the computer time and memory needed easily become too large.  Another
way of getting around this problem is due to Makse et al.\ 
\cite{Makse}. Here the (Fourier-space) filter function is modified by
the introduction of a large momentum cut-off through the use of a
modified Bessel function in the Fourier transform of the power
spectrum. They show that this large momentum cut-off, while irrelevant
for the large scale behavior in real space, is essential in order to
suppress the aliasing effect and thereby obtaining surfaces with the
desired scaling properties over the entire system size.

In this paper we report on an alternative filtering algorithm based on
wavelets which a priori, and without modifications, gives rise to
self-affine correlations which extends (up to finite size effects)
over the entire system independently of its size. This algorithm is
also computationally cheaper than the traditional (or modified)
Fourier filtering algorithm.

This paper is organized as follows: In Section \ref{Sec:2} we briefly
review the defining properties of self-affine surfaces.  Here we also
include some results which will prove useful to us later.
Section~\ref{Sec:3} is devoted to the outline of the new algorithm.
In Section \ref{Sec:4} we present numerical studies of the this
algorithm.  We conclude in Section \ref{Sec:5}.

\section{Self-affine surfaces}
\label{Sec:2}

We limit our discussion in this paper to $(1+1)$-dimensional surfaces, 
which we will call {\it profiles.\/}
A (statistically) self-affine profile, $h(x)$, is by definition a
structure which remains (statistically) invariant under the following
scaling relation\footnote{Strictly speaking, one should in order to
    fully define the self-affine structure, also introduce the
    topothesy which is defined as the length-scale $l$, over which the
    RMS-height, $\sigma$, is  $\sigma(l)=l$.  However, we will not
    explicitly need this latter quantity here, and will therefore
    simply neglect any further reference to it.}
\begin{mathletters}
    \label{SA-def}
\begin{eqnarray}
    x  &\rightarrow& \lambda x, \\
    h  &\rightarrow& \lambda^H h\;.
\end{eqnarray}
\end{mathletters}
Here $\lambda$ is a real number and $H$, known as the roughness or
Hurst exponent, characterizes this invariance. This exponent is usually
in the range from zero to one. When $H=1/2$, the profile is not
correlated.  An example of such a profile is the Brownian motion in
one dimension.  In this case, we interpret time as $x$ and $h(x)$ as
the position on the Brownian particle at time $x$. When $H>1/2$
the profile is persistent, while when $H<1/2$ it is anti persistent. 

We show in Fig.\ \ref{Fig:1} an example of a self-affine profile 
generated by the algorithm to be presented in this paper.   

From the scaling relation (\ref{SA-def}), one can often relatively
easily derive scaling relations for related quantities. In this paper we
will later explicitly need the scaling relation for the second order
structure function
\begin{eqnarray}
    S(\Delta x) &=& \left\langle \left| h(\Delta x + x)
    - h(x)\right|^2\right\rangle_x\;,
\end{eqnarray}
where $\langle\cdot\rangle_x$ represents the average over the position
variable $x$, and the power spectrum $P(q)$, defined as the Fourier
transform of the height-height correlation function.  They scale as
\cite{Feder,Meakin}
\begin{eqnarray}
    \label{struc-func}
    S(\Delta x) &\sim&  (\Delta x)^{2H}\;, 
\end{eqnarray}
and
\begin{eqnarray}
    \label{power-spec}
    P(q)        &\sim&  q^{-2H-1}\;.
\end{eqnarray}
We will make use of these two scaling relations in
Section~\ref{Sec:4}.

\section{The wavelet filtering algorithm}
\label{Sec:3}

Recently, the wavelet transform \cite{Press,Daubechies,Mallat98} has been used to
analyze self-affine profiles \cite{Sahimi,Simonsen}.  The idea behind this
analysis is as follows: We denote the wavelet transform of a
function $h(x)$ by ${\cal W}[h](a,b)$, 
where $a$ and $b$ are the scaling and location
variables respectively. They form together the wavelet domain. In
Ref.\ \ref{Ref:Simonsen} the authors introduced what they called the
average wavelet coefficient function, defined as $W[h](a) = \langle
  \left| {\cal W}[h](a,b) \right| \rangle_b$, where
$\langle\cdot\rangle_b$ denotes the average over all the location
parameters $b$ corresponding to one and the same scale $a$.  For a
self-affine function, $h(x)$, this quantity should scale as
\begin{eqnarray}
    \label{AWC-SA}
    W[h](a) &\sim&  a^{H+1/2}.
\end{eqnarray}

In much the same way as the Fourier filtering algorithm is used for
generating self-affine profiles via the fast Fourier transform, a
wavelet based filtering technique can be based on Eq.\ (\ref{AWC-SA}) in
combination with the fast wavelet transform. The output of the fast
wavelet transform \cite{Press,Daubechies,Mallat98} is a vector organized as a
collection of various levels or hierarchies all of different lengths
where each level, $\ell$, is associated with a corresponding scale
$a_\ell$.  The two first components of this vector, also known as
level $\ell=0$, are associated with the scaling function. All the
other components are ``true'' wavelet coefficients, such that at level
$\ell$, corresponding to scale $a_\ell=2^{-\ell}$, there are
$N_\ell=2^\ell$ coefficients. These coefficients (using our
convention) are arranged such that the coefficients of the highest
level are found at the end of the vector, and the levels decrease
monotonically towards the top of the vector, corresponding to the
level $\ell=0$.

Hence, the wavelet based algorithm which we will be referring to as the
{\it wavelet filtering algorithm\/} (WFA), consists of the following three
steps:
\begin{itemize}
\item Generate in the wavelet-domain  normalized uncorrelated
    Gaussian numbers $\{\eta_i\}$, with $i=1,2,\ldots,N$ where $N$ is
    the number of discrete points $x_i$ that together with $h_i=h(x_i)$
    constitute the self-affine profile.
    $N$ is assumed, due to the use of the fast wavelet
    transform, to be a power of $2$.
\item Filter these random numbers according to 
    \begin{eqnarray}
           w_i &=& \left( a_{\ell(i)} \right)^{H+1/2} \, 
                \frac{\eta_i}{\langle\left| \eta \right|\rangle_{\ell(i)}}\;,
           \qquad i=1,2,\ldots,N,\nonumber
    \end{eqnarray}
    to obtain the wavelet coefficients $\{w_i\}$.  Here
    $a_{\ell(i)}=2^{-\ell(i)}$ represents the scale, at level
    $\ell(i)$, where $\ell(i)$ is defined as the level corresponding
    to the location index $i$ of the vector $w_i$ (or $\eta_i$).
    Furthermore, $\langle\left| \eta \right|\rangle_{\ell(i)}$
    represents the average of the absolute value of those $\eta_i$ that
    together form level $\ell(i)$.
 \item Perform the inverse fast wavelet transform on $\{w_i\}$, with the
    (compactly supported) wavelet of your choice, to obtain the
    (real-space) self-affine profile of predefined Hurst exponent $H$.
\end{itemize}

With the wavelet filtering algorithm, good quality self-affine surfaces with
predefined Hurst exponent can be generated.  In Fig.\ \ref{Fig:1} we
show an example of a  self-affine surface of Hurst exponent $H=0.6$ 
and length $N=4096$ generated by the algorithm just outlined. It is 
worth noting that the above three steps can be modified in order to
deal with surfaces in higher dimensions \cite{ingve}. In this case the
speed of the surface generating algorithm becomes very important.

One of the prominent features of the wavelet transform is that the
basis functions, the wavelets, are localized in both space and
frequency. This has as a consequence, among others, that there is
no aliasing, or at least heavily suppressed as compared
to the Fourier transform. This implies that the wavelet filtering
algorithm should automatically result in surfaces which have the
desired correlations over the entire length of the profile, and not
just a small fraction of it.  Hence, independent of the
system size, the WFA is capable of generating profiles
with well-defined long-range correlations.  We will demonstrate the
validity of this claim in Section \ref{Sec:4}.

Before turning to the numerical studies of the WFA, we add some remarks 
regarding the computational efficiency of this algorithm. The most 
time-consuming part of the WFA is the
inverse wavelet transform. To a good approximation, at least for larger
system sizes, this time determines the overall computational time of
the entire algorithm. The fast wavelet transform needs ${\cal O}(cN)$
operations, where $c$ is a positive real number which value depends on
the wavelet used \cite{Mallat98,Mallat}.
Thus, the number of operations need for generating a
surface by WFA is {\it linear\/} in the number of points belonging to the
profile. In comparison, the Fourier filtering algorithm, which speed
is mainly controlled by the fast Fourier transform, needs ${\cal
O}(N\log_2N)$ to generate a profile.  For large system sizes the
difference in execution time between WFA and the Fourier filtering
algorithm becomes significant.

\section{Numerical results}
\label{Sec:4}

In order to test numerically the predictions of the previous section,
we have chosen to study the second order structure function, $S(\Delta
x)$, and the power spectrum $P(q)$ of self-affine profiles generated
with the wavelet filtering algorithm. The appropriate scaling relations
for these two quantities are given by Eqs.\ (\ref{struc-func}) and
(\ref{power-spec}). They will provide us with independent
information enabling us to accurately quantify over which length
scales the self-affine correlations exist. The numerical experiments,
for which the results will be presented shortly, were performed as
follows: We generated, by WFA, an ensemble of long self-affine
profiles all with the same Hurst exponent $H$.  For each profile the
structure function and power spectrum were calculated, and these
were averaged over the ensemble of profiles.

In Fig.\ \ref{Fig:2} we give the numerical results for the second
order structure function obtained as described above. The predefined
Hurst exponents, used by the surface generator, were from bottom to
top $H=0.8$, $0.6$, $0.4$, and $0.2$ as indicated in the figure.  The
length of each profile was $N=2^{25}= 33\, 554\, 432$. The number of
profiles used in obtaining the averages was $N_h=50$, and the wavelet
used was of the Daubechies-type ($D12$)
\cite{Press,Daubechies,Mallat98}.  The dashed lines are regression fits
to the numerical data. They corresponds (from bottom to top) to Hurst
exponents of $H=0.80\pm 0.01$, $0.60\pm 0.01$, $0.41\pm 0.01$, and
$0.22\pm 0.02$, all consistent, within the errorbars, with the
predefined exponents given above.  One easily observes from Fig.\ 
\ref{Fig:2} that the correlations extend over all scales except for
the largest lags $\Delta x$.  The reason that the last few large lags
do not fit into this general picture is due to finite size effects. We
have also undertaken the above analysis for different types of
wavelets, taken from the Daubechies family, and for various system
sizes, finding no results that are inconsistent with those presented
in Fig.\ \ref{Fig:2}.

In order to make the comparison with the Fourier filtering algorithm
even more apparent, we present in Fig.\ \ref{Fig:3} the average power
spectrum obtained using the same surfaces as were used in Fig.\ 
\ref{Fig:2}. The correlations again span most scales. The dashed
regression fits lead to the following exponents~(from bottom to top):
$H=0.80\pm 0.01$, $0.61\pm 0.01$, $0.41\pm 0.01$, and $0.20\pm 0.01$,
which again is in excellent agreement with the values of the Hurst
exponent used for the generation of the underlying profiles.

Figs.\ \ref{Fig:2} and \ref{Fig:3} indicate that the self-affine 
correlations span all but the largest scales of the profiles. We
stress that this is a generic property of the wavelet filtering
algorithm, and no modification of the algorithm is needed in order to handle
large system sizes in a satisfactory manner. This is a consequence of
the celebrated property of the wavelets being localized {\it both\/} in
space and frequency.

The calculations of this paper were performed on a SGI/Cray Origin 2000
supercomputer based on the R10000 chip from SGI. On this machine the
average {\sc cpu} time needed for generating a profile of the length used
above ($N=2^{25}$) was $t_{\mbox{{\sc wfa}}}=45$s and $t_{\mbox{{\sc
            ffm}}}=125$s for the WFA and the traditional (or
modified) Fourier filtering algorithm respectively. Hence the speedup
gained by using the wavelet filtering algorithm over the Fourier filter
algorithm is close to a factor $3$.  For system sizes $N\sim 10^3$, we
could not observe any significant different between the two algorithms.

\section{Conclusions}
\label{Sec:5}

To summarize, we have introduced a fast and simple algorithm for generating
long (or short) self-affine profiles. This algorithm, named the wavelet
filtering algorithm, is demonstrated to overcome the problem related to
the aliasing effect which the traditional Fourier filtering algorithm is
troubled with.  Furthermore, the wavelet based filtering technique
outperforms its Fourier-domain counterpart by large margins with respect 
to computational costs, at least for large system sizes.

\acknowledgements

I.S.\ would like to thank the Research Council of Norway and Norsk
Hydro ASA for financial support.  A.H.\ thanks H.N.\ Nazareno and
F.A.\ Oliveira for warm hospitality and the I.C.C.M.P.\ for support.
This work has received support from the Research Council of Norway
(Program for Supercomputing) through a grant of computing time.


\end{multicols}

\newpage

\widetext

\begin{figure}
    \caption{A self-affine profile, $h(x)$, of length $N=4096$ and Hurst
        exponent $H=0.6$ generated by the wavelet filtering
        algorithm. The wavelet used in order to generate the profile was
        the D12 Daubechies wavelet.}
    \label{Fig:1}
\end{figure}

\begin{figure}
    \caption{The average second order structure function, $S(\Delta x)$, 
      obtained by averaging over $N_h=50$ samples (for given $H$) of
      the self-affine profile $h(x)$ generated by WFA. All profiles
      were of length $N=2^{25}=33\,554\,432$. The Hurst exponents used
      were (from bottom to top) $H=0.8$, $0.6$, $0.4$, and $0.2$, as
      indicated in the figure.  The dashed lines are the best
      regression fits corresponding respectively to Hurst exponents
      (from bottom to top) $H=0.80\pm 0.01$, $0.60\pm 0.01$, $0.41\pm
      0.01$, and $0.22\pm 0.02$.}
    \label{Fig:2}
\end{figure}

\begin{figure}
    \caption{The average power spectrum $P(q)$
      obtained by averaging over $N_h=50$ samples (for given $H$) of
      the self-affine profile $h(x)$ generated by WFA. All profiles
      were of length $N=2^{25}=33\,554\,432$. The Hurst exponents used
      were (from bottom to top) $H=0.8$, $0.6$, $0.4$, and $0.2$, as
      indicated in the figure.  The dashed lines are the best
      regression fits corresponding respectively to Hurst exponents
      (from bottom to top) $H=0.80\pm 0.01$, $0.61\pm 0.01$, $0.41\pm
      0.01$, and $0.20\pm 0.01$.}
    \label{Fig:3}
\end{figure}

\setcounter{figure}{1}
\newcommand{\mycaption}[2]{\begin{center}{\bf Figure \thefigure}\\{#1}\\{\em #2}\end{center}\addtocounter{figure}{1}}
\newcommand{\myauthor}{I. Simonsen and A. Hansen}
\newcommand{\mytitle}{A new fast algorithm for generating long self-affine profiles}


\newpage
\begin{figure}
    \begin{center}
            \epsfig{file=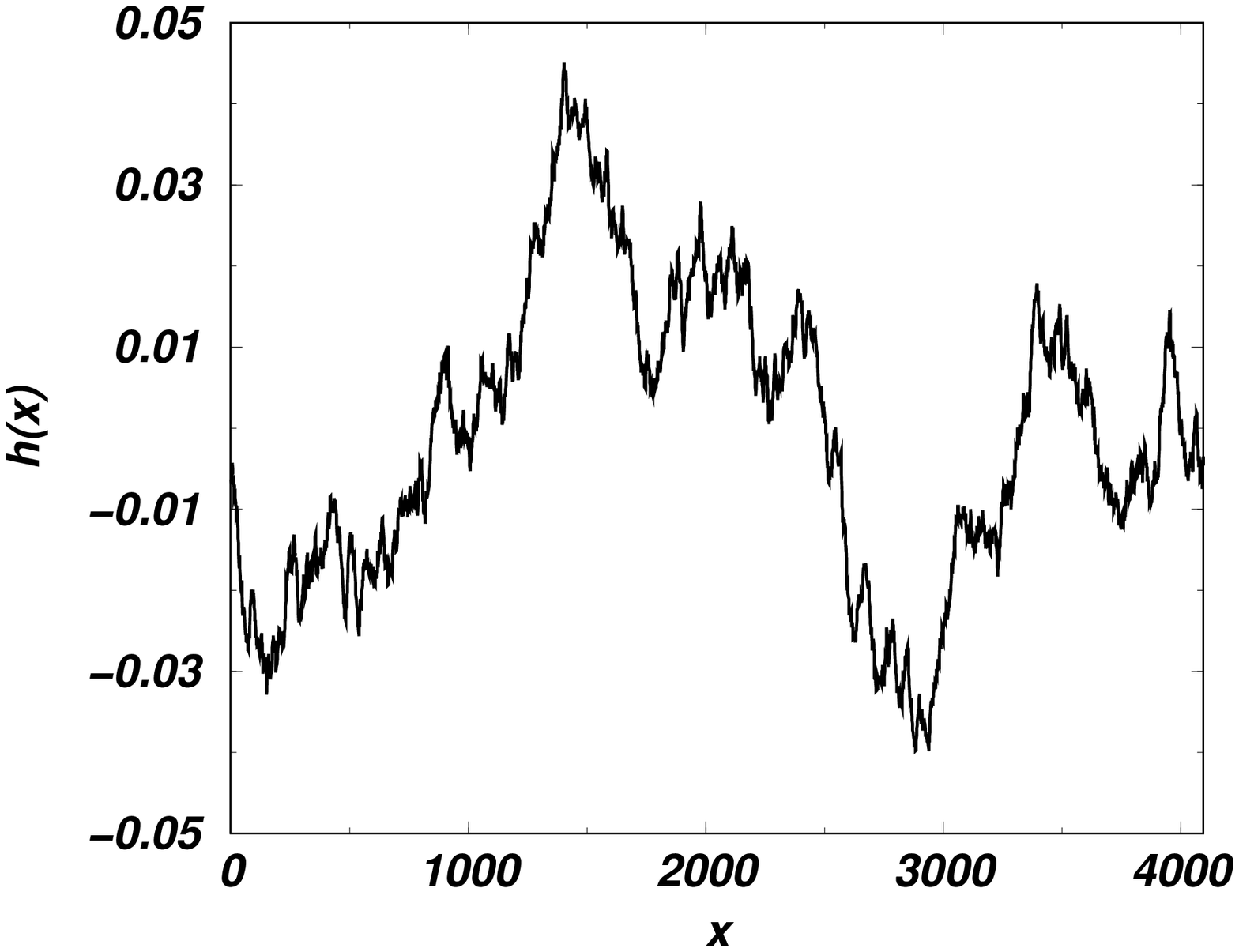,width=12.5cm,height=8.5cm} 
    \end{center}
    \mycaption{\myauthor}{\mytitle}
\end{figure}

\newpage
\begin{figure}
    \begin{center}
            \epsfig{file=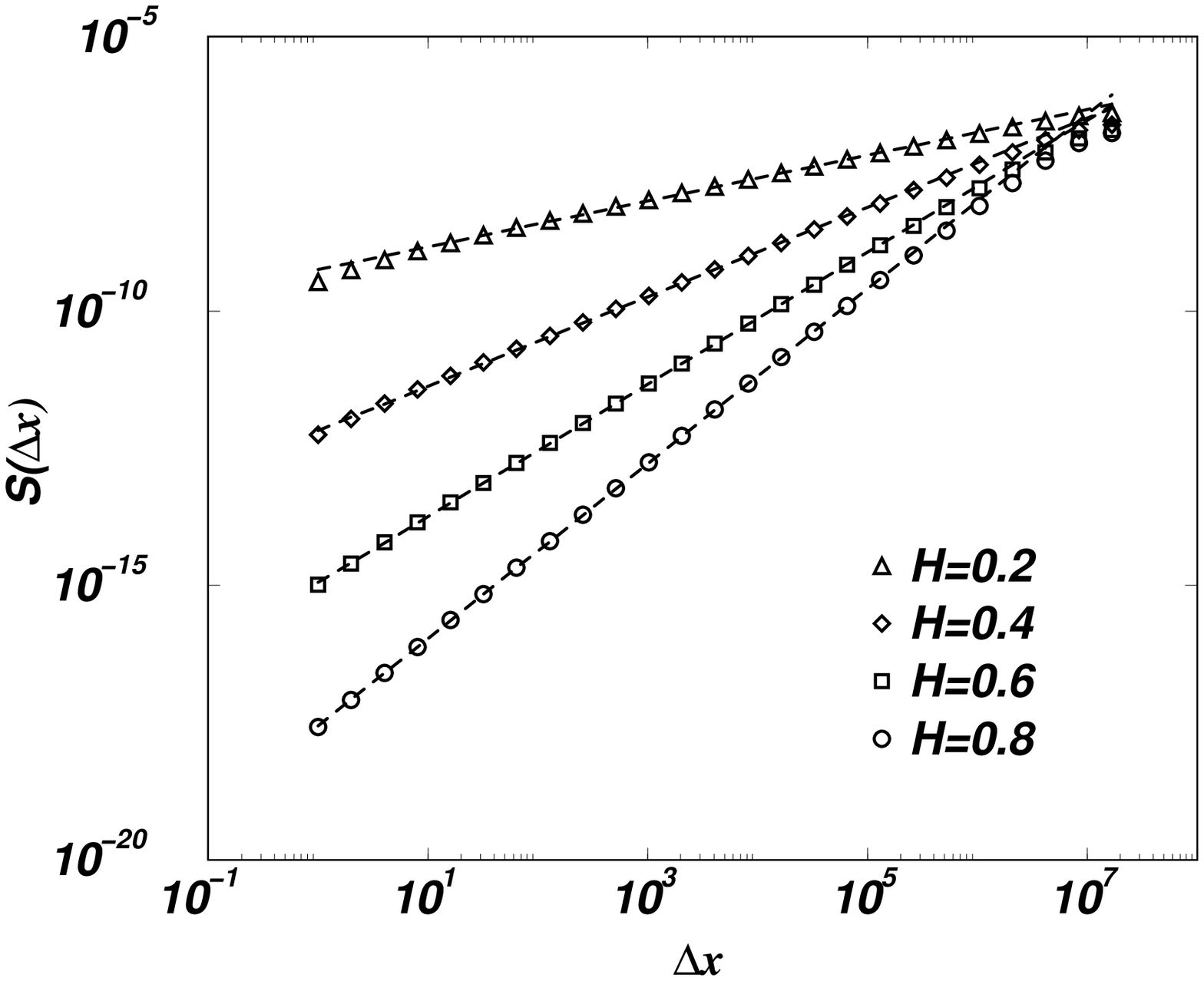,width=12.5cm,height=12.5cm} 
    \end{center}
    \mycaption{\myauthor}{\mytitle}
\end{figure}

\newpage
\begin{figure}
    \begin{center}
            \epsfig{file=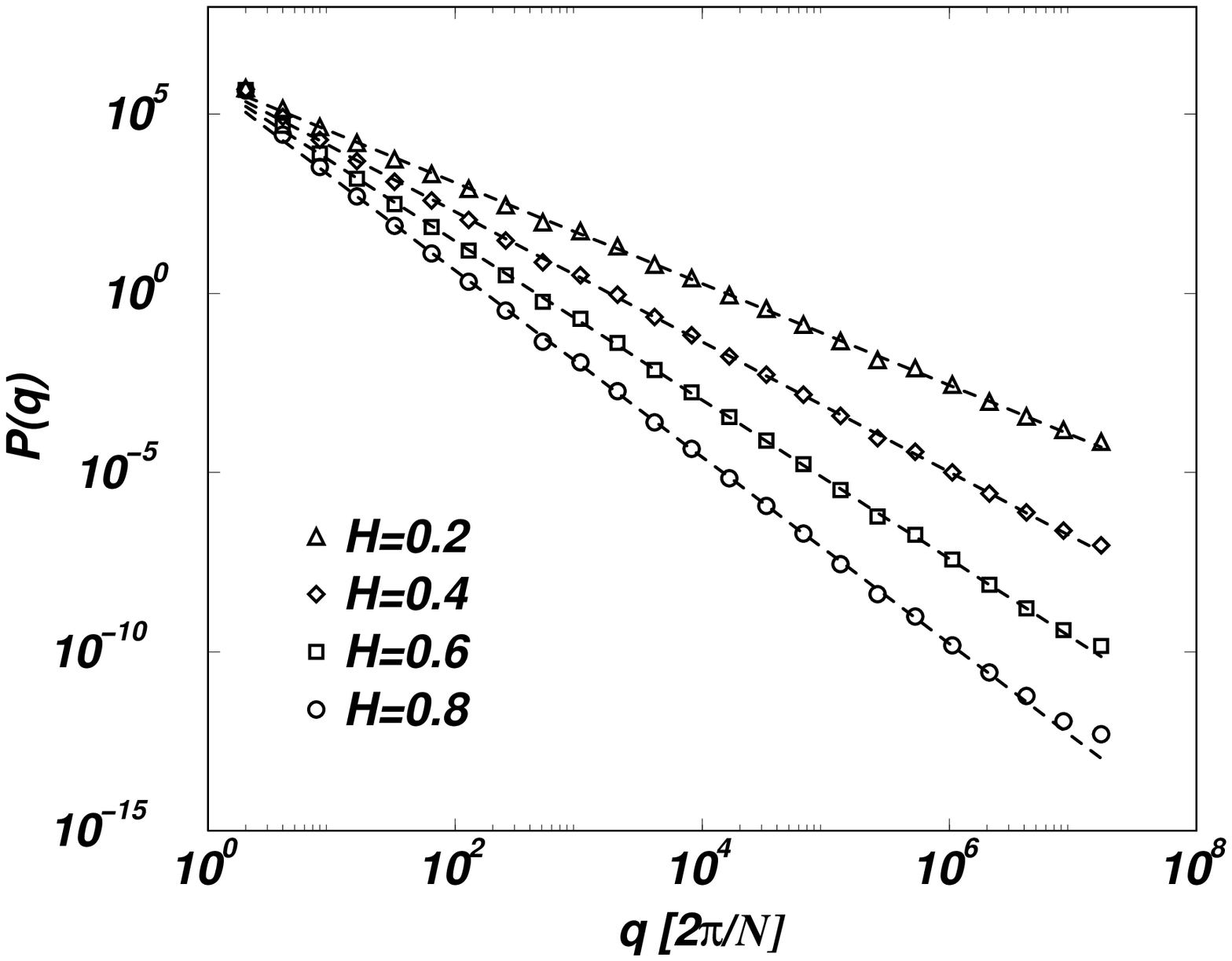,width=12.5cm,height=12.5cm} 
    \end{center}
    \mycaption{\myauthor}{\mytitle}
\end{figure}

\end{document}